\documentclass[11pt,draftcls,onecolumn]{IEEEtran}
\usepackage{graphicx}
\usepackage{cite}
\usepackage{amsmath,amsfonts}

\usepackage[boxed]{algorithm2e}
\title{Blind Denoising with Random Greedy Pursuits}

\author{
\IEEEauthorblockN{Manuel Moussallam\IEEEauthorrefmark{1},
Alexandre Gramfort\IEEEauthorrefmark{2}},
\IEEEauthorblockN{Laurent Daudet\IEEEauthorrefmark{1}, and Ga\"el Richard\IEEEauthorrefmark{2}}
\\
\IEEEauthorblockA{\IEEEauthorrefmark{1}Institut Langevin - Univ. Paris
Diderot - ESPCI ParisTech, first.last@espci.fr, 1 rue Jussieu 75005 Paris, France}\\
\IEEEauthorblockA{\IEEEauthorrefmark{2}Institut Mines-Telecom - Telecom
ParisTech - CNRS/LTCI, first.last@telecom-paristech.fr \\ 37-39 rue Dareau, 75014 Paris, France}
}

\begin{document}
\maketitle

\begin{abstract}

Denoising methods require some assumptions about the signal of interest and the noise. While most denoising procedures require some knowledge about the noise level, which may be unknown in practice, here we assume that the signal expansion in a given dictionary has a distribution that is more heavy-tailed than the noise. We show how this hypothesis leads to a stopping criterion for greedy pursuit algorithms which is independent from the noise level. Inspired by the success of ensemble methods in machine learning, we propose a strategy to reduce the variance of greedy estimates by averaging pursuits obtained from randomly subsampled dictionaries. We call this denoising procedure Blind Random Pursuit Denoising (BIRD). We offer a generalization to multidimensional signals, with a structured sparse model (S-BIRD). The relevance of this approach is demonstrated on synthetic and experimental MEG signals where, without any parameter tuning, BIRD outperforms state-of-the-art algorithms even when they are informed by the noise level. Code is available to reproduce all experiments.

\end{abstract}
\vspace{-2mm}
\begin{center} \bfseries EDICS Category: DSP-SPARSE \end{center}
\vspace{-5mm}
\section{Introduction}
Time series obtained from experimental measurements are always contaminated by noise. Separating the informative signal from the noise in such raw data is called \emph{denoising} and requires some assumptions on the signals and/or noise, for instance imposing a sparse model on the discrete signal $y$, of finite size $N$: $$y = \Phi \alpha + w \enspace ,$$
where $\Phi\in \mathbb{R}^{N \times M}$ is a (usually overcomplete) dictionary of $M$ elementary objects $\phi_m$ called atoms and assumed normalized (\emph{i.e.} $\forall m, \|\phi_m\|_2=1$), $\alpha$ is a sparse vector (\emph{i.e.} $\|\alpha\|_0 = k \ll M$), and $w$ is the additive noise to be removed. This model thus expresses the informative part of the signal as a sparse expansion in $\Phi$ and implicitly states that the noise component has no such expansion. Under this assumption, a denoised estimate $\hat{y} = \Phi \hat{\alpha}$ of $y$ can be obtained by solving:
\begin{equation}
\hat{\alpha} = \arg \min_{\alpha \in \mathbb{R}^{M}} \|\alpha\|_0 \text{ subject to } \|y - \Phi \alpha \|_2 \leq \epsilon \enspace.
\label{l0_problem}
\end{equation}
The value $\epsilon$ must be chosen according
to the noise level (\emph{i.e.} the norm of $w$).
As problem \eqref{l0_problem} is NP hard, it is approximately solved using greedy
algorithms~\cite{mallat-zhang:93,Bertrand1994,Durka1995,quiroga-etal:03,Durka2005}, or via convex relaxations (\emph{e.g.} Basis Pursuit
Denoising~\cite{Chen1998}). A greedy algorithm,
such as Matching Pursuit (MP)~\cite{mallat-zhang:93} and variants, will iteratively build an
estimate $\hat{y}$ by selecting atoms in
$\Phi$ and updating a residual signal accordingly.
This latter class of methods suffers from two main limitations: (i) choosing a good value for $\epsilon$, \emph{i.e.} in practice a stopping rule for the algorithm,  requires some knowledge on the noise variance, and (ii) the obtained approximation strongly depends on the dictionary design.


The contributions of this paper are threefold. First, we derive a data-driven stopping criterion for greedy pursuits based on order statistics. This technique allows denoising without knowledge of the noise variance. Second, we show how randomized greedy pursuits can be combined to improve the performance and reduce the dependency on the dictionary choice. This new algorithm, called BIRD, can be generalized to the case of multidimensional signals (S-BIRD). Third, we use popular synthetic signals to compare the performance of the proposed method with state-of-the-art techniques (soft and hard thresholding using cycle spinning~\cite{Coifman1995}, stochastic MP~\cite{Durka2001} and randomized MP~\cite{Elad2009}). Results on experimental data obtained with magnetoencephalography (MEG) are presented.

\section{Blind Denoising with Randomized Pursuit}
\subsection{Stopping Criterion for Greedy Denoising Methods}
Greedy algorithms require a stopping criterion to control model complexity, and avoid under- or over-fitting. For denoising purposes, this stop ideally occurs when the residual equals the noise and all atoms selected so far only explain the signal. In practice, a clear distinction between signal atoms and noise atoms is not always available. In this context, an interesting measure with greedy approaches is the \emph{normalized coherence} of a signal $y$ in $\Phi$ as defined in~\cite{mallat-zhang:93}:
\begin{equation}
\lambda_{\Phi}(y)=\sup_{\phi\in\Phi}\frac{|\langle y,\phi\rangle|}{\|y\|_2} \enspace .
\end{equation}
Let $r^{n}$ be the residual signal at iteration $n$, it's energy decay can be expressed in terms of the normalized coherence by:
\begin{equation}
\frac{\|r^{n}\|_2^{2}}{\|r^{n-1}\|_2^{2}}	=1-\lambda_{\Phi}^{2}(r^{n-1}) \enspace .
\label{decay_to_lambda}
\end{equation}
This relation is essentially used to bound from above the convergence of the algorithm using the \emph{coherence} of the dictionary $\Lambda(\Phi)	=	\inf_{x\in\mathbb{R}^{N}}\left(\lambda_{\Phi}(x)\right)$. This value is useful to describe the worst case convergence scenario, \emph{i.e.} the convergence rate for the signal that is least correlated with $\Phi$. Considering the noise signal $w$ as a realization of a stochastic process, one may be also interested in the value:
\begin{equation}
\Lambda_W(\Phi)	=	\mathbb{E}\left[\lambda_{\Phi}(w)\right] \enspace .
\label{coherence_noise}
\end{equation}
Denoising can then be achieved by selecting only atoms whose normalized coherence is significantly higher than this value \cite{Durka1995,mallat-zhang:93}. Estimating \eqref{coherence_noise} is however uneasy, and is typically learned from a training set \cite{Durka1995}. 

The novelty of our approach is to propose a closed form estimate of $\Lambda_W(\Phi)$ based on a stochastic argument and order statistics. Let us consider the projections of $w$ over $\Phi$ as $M$ realizations $z_m^{w}$ of a random variable (RV) $Z^{w}$:
\begin{equation}
\forall m \in [1,M], z_m^{w} = \frac{|\langle w,\phi_m\rangle|}{\|w\|_2} \enspace .
\label{z_w_projscore}
\end{equation}
In most cases, $M$ is greater than $N$. It implies that the $z_m$ are
not independent: their joint distribution is intricate. However, for
analytical simplifications, we will assume they are i.i.d. This
simplification will allow us to derive a new bound that turns out to
be near-optimal in the proposed experimental framework.
When run on $w$ (\emph{i.e.} pure noise), 
a greedy algorithm such as MP, or Orthogonal MP (OMP \cite{Pati1993}), will typically select the atom that maximizes \eqref{z_w_projscore}. Let us denote by $Z_{(M)}^{w}$ the RV describing the maximum projection value among $M$ samples of $Z^{w}$. It is also known as the last order statistic of $Z^{w}$, and its cumulative density function writes (see for instance \cite{Wilks1947}):
\begin{equation}
F_{(M)}^{Z^{w}}(z) = M \int_{0}^{z} \left( F_Z^{w}(z')^{M-1} f_Z^{w}(z')\right) dz' \enspace ,
\label{eq:max_cdf}
\end{equation}
where $f_Z^{w}$ (respectively $F_Z^{w}$) is the probability (respectively cumulative) density function, PDF (respectively CDF) of $Z^{w}$.
Given an assumed i.i.d. distribution of the noise projections in a dictionary of size $M$, \eqref{eq:max_cdf} gives a closed form formula for the CDF of the maximum.

The intuition behind this work writes as follows: the value $p=1-F_{(M)}^{Z^{w}}(z)$ is the probability that the maximum correlation between a dictionary element and a pure noise is to be greater than $z$. Thus we need to design a dictionary such that unlikely observations indicate the presence of a signal.

Let us now make the assumption that the dictionary is designed such that: (i) the projections of the noise on its atoms are distributed according to a zero mean Gaussian distribution (GD) and (ii) the distribution of the projections of the informative part has a heavier tail than the GD. The GD model fits well a variety of practical situations (\emph{e.g.} white noise in a windowed-Fourier dictionary) and is more general than the standard Gaussian noise hypothesis.
A reasonable model for $Z^{w}$ is thus a half-normal RV, for which \eqref{eq:max_cdf} is easily computed. This allows us to replace the value in \eqref{coherence_noise} by:
\begin{equation}
\Lambda_W(\Phi, p) = \frac{\sqrt{2}}{N}\sqrt{\left(1-\frac{2}{\pi}  \right) }\mbox{erfinv}\left((1-p)^{\frac{1}{M}}\right) \enspace ,
\label{eq:boundary}
\end{equation}where $\mbox{erfinv}$ is the inverse error function. 

The parameter $p$ expresses the confidence in the model and thus controls how much an approximation shall fit the data. Large values of $p$ can lead to overfitting while small values can be too conservative. In this sense, this parameter plays a similar role to the more classical approximation error in \cite{mallat-zhang:93}. However, it is important to emphasize that $p$ is set for a given dictionary independently of the noise level.
Experimental results testing the sensibility of the method with respect to $p$ are given in supplementary material.

\subsection{Double Randomization}
The underlying assumption of \eqref{l0_problem} is that the \textit{sparsest} representation is the optimal choice. However, as shown by Elad et al \cite{Elad2009}, a better strategy (in the sense of the mean squared error) is to sample a set of $J$ random sparse approximations $\{\hat{y}^j\}_{j=1..J}$ and average them.
Such an approach would be named \emph{ensemble method} in the statistical learning literature
(see \emph{e.g.} \cite{esl} chap.~16).
The $J$ randomized greedy decompositions are run in parallel with the following probabilistic selection
procedure. Let $r^{n-1}$ be the residual signal at iteration
$n$. The $n$-th element index $\gamma^n$ to be selected is chosen at random among the $M$
columns $\phi_m$ of $\Phi$ with greater probabilities, \emph{i.e.} with large inner products
$|\langle \phi_{\gamma^n}, r^{n-1}\rangle|$.

Our strategy, based on the work in~\cite{Moussallam2012}, extends this
idea with a \emph{Random Forest} -like approach~\cite{Breiman2001}.
At each iteration, the most correlated element from a random subset
$\Phi_n \subset \Phi$ is selected. Here, $\Phi_n$ will be about 50 times smaller than $\Phi$.
This spares us the computation of the
$M$ inner products, while proper randomization scheme allows us to browse the
whole dictionary across iterations, a strategy particularly interesting when
dictionary elements are finely located in time and frequency (see
\cite{Moussallam2012}). Running $J$ instances of this pursuit on a random sequence of subdictionaries (\emph{i.e.} the equivalent of $J$ random trees) yields a set $\{\hat{y}^j\}_{j=1..J}$ of sparse approximations. They can then be averaged in order to obtain the denoised signal: $\tilde{y} = \frac{1}{J} \sum_j \hat{y}^j$.
The complete algorithm is detailed in Algorithm~\ref{algo:bird}.

\begin{algorithm}
\KwIn{$y$,$\Phi$, $J$, $\Lambda_W(\Phi,p)$}
\KwOut{$\tilde{y}$}
\For{$j=1..J$}{
initialization: $n=0$,$r^0=y,\hat{y}^j=0$\;
\While{condition}{
$n \leftarrow n+1$\;
Draw at random $\Phi_n \subset \Phi$\;
Select $\phi_{\gamma^n}=\arg\max_{\phi \in \Phi_n} |\langle r^{n-1}, \phi\rangle|^{2}$\;  
Update $\hat{y}^j \leftarrow \hat{y}^j +  \langle r^{n-1}, \phi_{\gamma^n}\rangle \phi_{\gamma^n}$\;
and $r^n = y - \hat{y}^j$\;
condition  = $\lambda_\Phi (r^{n-1}) > \Lambda_W(\Phi,p)$\;
}
}
$\tilde{y} = \frac{1}{J} \sum_j \hat{y}^j$\;
\caption{Blind Random Pursuit Denoising (BIRD)}
\label{algo:bird}
\end{algorithm}

\section{Structured Sparse model}
In case of data acquired with multiple sensors, the sparse model can be extended to take the structure
of the data into account. Let $Y \in \mathbb{R}^{N \times C}$ be the data matrix formed by stacking the signals recorded by $C$ sensors.
Given the same dictionary $\Phi$, one seeks an approximate $\hat{Y}=\Phi \hat{A}$ of $Y$
as a sparse expansion in $\Phi$. The unstructured problem reads:
\begin{equation}
\hat{A} = \arg \min_{A \in \mathbb{R}^{M\times C}} \|A\|_{0} \text{ subject to } \|Y - \Phi A \|_F \leq \epsilon \enspace ,
\label{l0_matrix_problem}
\end{equation}
where $\|.\|_F$ stands for the Frobenius norm and  $\|.\|_0$ is the number of non-zero entries.
The signal model for one sensor reads:
\begin{equation}
 y_c = \Phi(\alpha_c \odot s) + w_c \enspace ,
\end{equation}
where $y_c$ is the $c$-th column of $Y$, $w_c$ is the noise recorded by sensor $c$, $s$ is a binary sparse vector of zeroes or ones independent of the sensor, and $\alpha_c$ is a weight vector specific to the sensor that has the same support as $s$ and whose values are all zeroes if $c$ is not in the set $\Gamma_C$. The notation $\odot$ stands for the element-wise multiplication.
In a matrix form, this writes $Y = \Phi A + W$ with $A$ a matrix whose $c$-th column is full of zeroes if $c$ is not in $\Gamma_C$ and whose $m$-th row is full of zeroes if $s[m]=0$. Such matrices typically arise when using mixed-norms for group-sparse approximation problems~\cite{kowalski08MN}, such as:
 \begin{equation}
 \hat{A} = \arg \min_{A \in \mathbb{R}^{M\times C}} \|A\|_{2,1} \text{ subject to } \|Y - \Phi A \|_F^{2} \leq \epsilon
 \label{lpq_problem}
 \end{equation}
 where $\|.\|_{2,1}$ is the $\ell_{2,1}$ norm as in \cite{kowalski08MN}.
Such problems are commonly addressed with greedy algorithms \cite{Tropp2005} or group soft-thresholding (Group-LASSO)\cite{Roth2008}. It is known as a Multiple Measurement Vector (MMV) problem in the signal processing literature \cite{Tropp2005}.

Adapting Algorithm~\ref{algo:bird} to the structured case requires a refined selection rule.
Typically an atom is selected if it maximizes the sum of the projections over all sensors~\cite{Tropp2005}.
However, only a fraction of the sensors may simultaneously record the signal from a source. Let $l$, $0 < l \leq 1$, be this fraction. Let $r_c^{n-1}$ be the residual signal of sensor $c$ and let us write $p_{c,m}^{n-1} = |\langle \phi_m , r_{c}^{n-1} \rangle|^2$ and $p_{(c),m}^{n-1}$ the ordered projections of $r_c^{n-1}$ on $\phi_m$ (\emph{i.e.} $p_{(1),m}^{n-1} \leq ... \leq p_{(C),m}^{n-1}$). Let us denote:
\begin{equation}
\phi^n = \arg \max_{\phi \in \Phi} \frac{1}{\lfloor lC \rfloor} \sum_{c=\lfloor C(1-l) \rfloor}^C p_{(c),m}^{n-1} \enspace .
\label{select_crit_maxofk}
\end{equation}
The selection procedure also yields a list $\Gamma_c^{n}$ of $\lfloor lC\rfloor$ sensors containing the atom, and where an update is necessary.

In theory, it is again possible to use order statistics to model the sum in (\ref{select_crit_maxofk}) and its maximum as RVs. In practice, 
a simple idea is to stop the decomposition once a given proportion of the most energetic signals have been denoised:
\begin{equation}
Cond \left[\lambda_\Phi (r_c^{n-1}) \right] = \frac{1}{\lfloor lC \rfloor} \sum_{c=\lfloor C(1-l) \rfloor}^C  \lambda_\Phi (r_{(c)}^{n-1}) > \Lambda_W(\Phi, p) \enspace ,
\label{mean_of_k_stop_crit}
\end{equation}
where the set $\{ \lambda_\Phi (r_{(c)}^{n-1}) \}$ corresponds to the $\lfloor lC \rfloor$ biggest values of $ \lambda_\Phi (r_{c}^{n-1})$. This criterion is coherent with the selection rule (\ref{select_crit_maxofk}). The complete procedure is summarized in Algorithm~\ref{algo:SBIRD}.
\begin{algorithm}
\KwIn{$Y$, $\Phi$, $J$,  $l$, $\Lambda_W(\Phi, p)$}
\KwOut{$\tilde{Y}$}
\For{$j=1..J$}{
initialization: $n=0$,$\forall c, r_c^0=y_c, \hat{y}_c^j=0$\;
\While{condition}{
$n \gets n+1$\;
Draw at random $\Phi_n \subset \Phi$\;
Select $\left( \phi_{\gamma^n},\Gamma_c^{n} \right) =\arg \max \frac{1}{\lfloor lC \rfloor} \sum_{c=\lfloor C(1-l) \rfloor}^C p_{(c),m}^{n-1}$\;
Update $\forall c \in \Gamma_c^{n}, \hat{y}_c^{j} \leftarrow \hat{y}_c^{j} +  \langle r_c^{n-1}, \phi_{\gamma^n}\rangle \phi_{\gamma^n}$\;
and $r_c^n = y_c - \hat{y}_c^{j}$\;
condition  = $Cond \left[\lambda_\Phi (r_c^{n-1}) \right]$\;
}
}
$\tilde{Y} = \frac{1}{J} \sum_j \hat{Y}^{j}$\;
\caption{Structured Blind Random Pursuit Denoising (S-BIRD)}
\label{algo:SBIRD}
\end{algorithm}
\vspace{-4mm}
\section{Experimental validation}
All the experiments and results shown in this section as well as those presented in the provided supplementary material can be reproduced using our Python code freely available online\footnote{http://manuel.moussallam.net/birdcode}.
\subsection{Synthetic examples}
A first set of experiments compares the proposed algorithm BIRD to existing mono-channel denoising techniques on synthetic signals and simulated MEG data. 
We present the performance of BIRD compared to various state-of-the art methods among which:
\begin{itemize}
\item Wavelet Shrinkage (WaveShrink) methods (with both soft and hard thresholding) using Daubechies wavelets and a Short-Time Fourier Transform (STFT) dictionaries. These dictionaries can be made shift-invariant using the Cycle-Spinning method~\cite{Coifman1995,Kamilov2012}.
\item Stochastic MP (SMP) as introduced in \cite{Durka2001}. In this method, each of the $J$ runs is performed on a subdictionary $\Phi_j \subset \Phi$, that is chosen at random once for each run, and kept unchanged in the whole decomposition.
\item Randomized OMP (RandOMP) as introduced in \cite{Elad2009}. In this method, atoms are selected at random in the complete dictionary $\Phi$ at every iteration of the $J$ runs.
\end{itemize}
These methods require a stopping criterion, typically set by fixing the reconstruction error in accordance with the noise level. In contrast, our algorithms select an atom in a random subdictionary at each iteration of the $J$ runs and derive their stopping criterion from the statistics of the projections as explained above.
For comparison, we present the results obtained by WaveShrink (respectively SMP and RandOMP) methods in an \textit{Oracle} case, that is when the true signal $y$ is known and used to set the target reconstruction errors in order to minimize the errors.

For all greedy approaches, we use an overcomplete dictionary $\Phi$ built as a union of Modulated Discrete Cosine Transforms (MDCT) of 6 different scales. For each basis, atoms are further replicated and shifted so as to form a highly overcomplete, shift-invariant, dictionary $\Phi$ of size $M\gg N$. We set the overfitting probability to $p=10^{-6} \sim 1/M$. This may seem a conservative value, but in practice the algorithm is not very sensitive to $p$ (See Fig.~3 in supplementary materials).

One can verify in Figure \ref{fig:single_synthetic} that BIRD has some advantages over alternative methods. Two observations can be made.
First, the double randomization scheme is valuable: with a limited number of runs $J$ the resulting denoised signal with BIRD presents less disturbing artifacts than SMP or RandOMP methods.
Second, the self-stopping criterion yields satisfying signal estimates in most cases. Note that no parameter is modified when varying the SNR. Given pure white noise as input, BIRD does not select any spurious atom. Given a pure sparse signal, BIRD will select atoms up to a very high reconstruction fidelity.
\begin{figure}
\centering
\includegraphics[width=8cm]{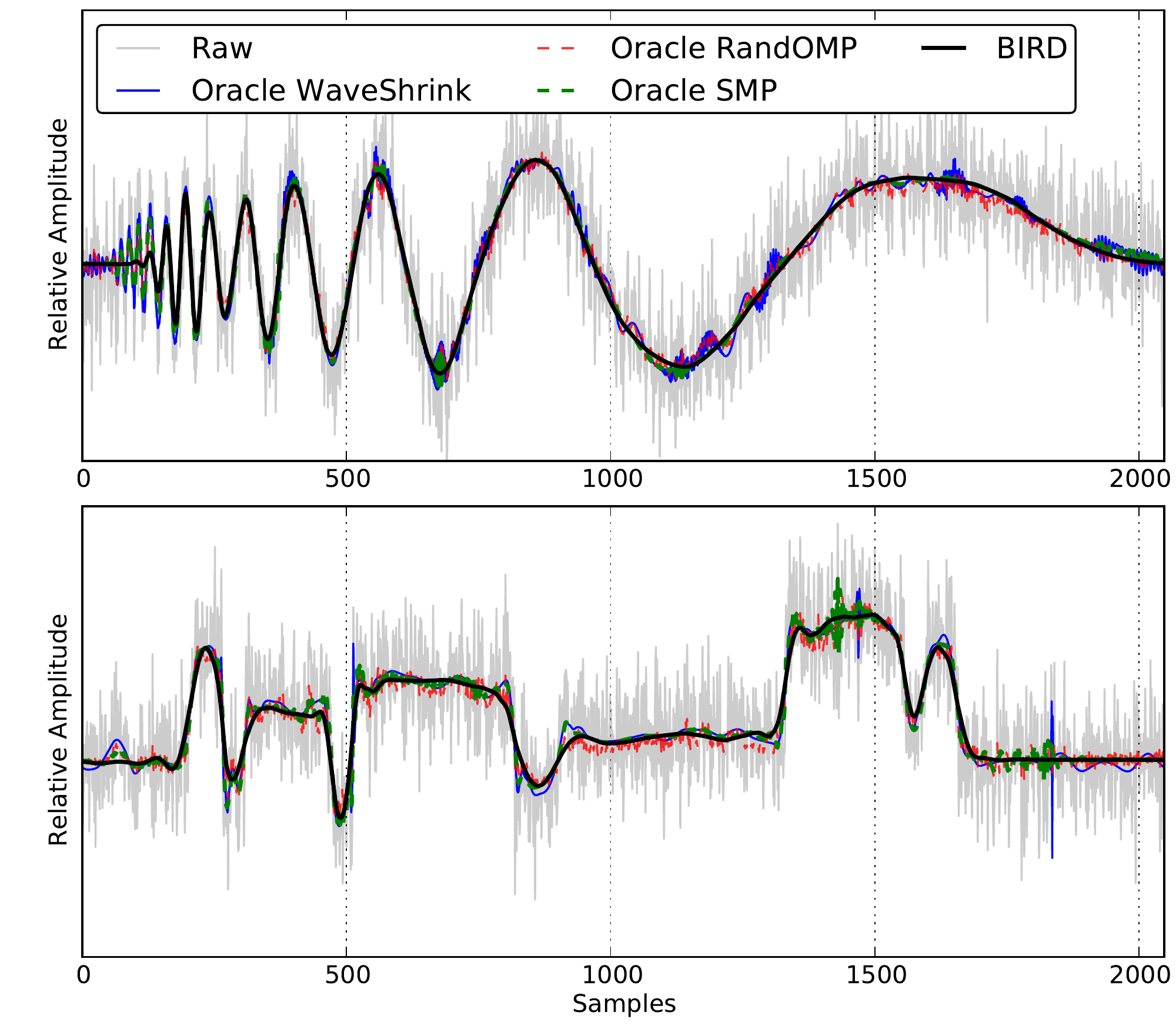}
\caption{Examples of denoising for synthetic signals \textit{Doppler} and \textit{Blocks}. SMP, RandOMP and BIRD are set with $J=30$ runs and use a multiscale MDCT dictionary.}
\label{fig:single_synthetic}
\end{figure}
\subsection{Simulated MEG data}
\begin{figure}
\centering
\includegraphics[width=8cm]{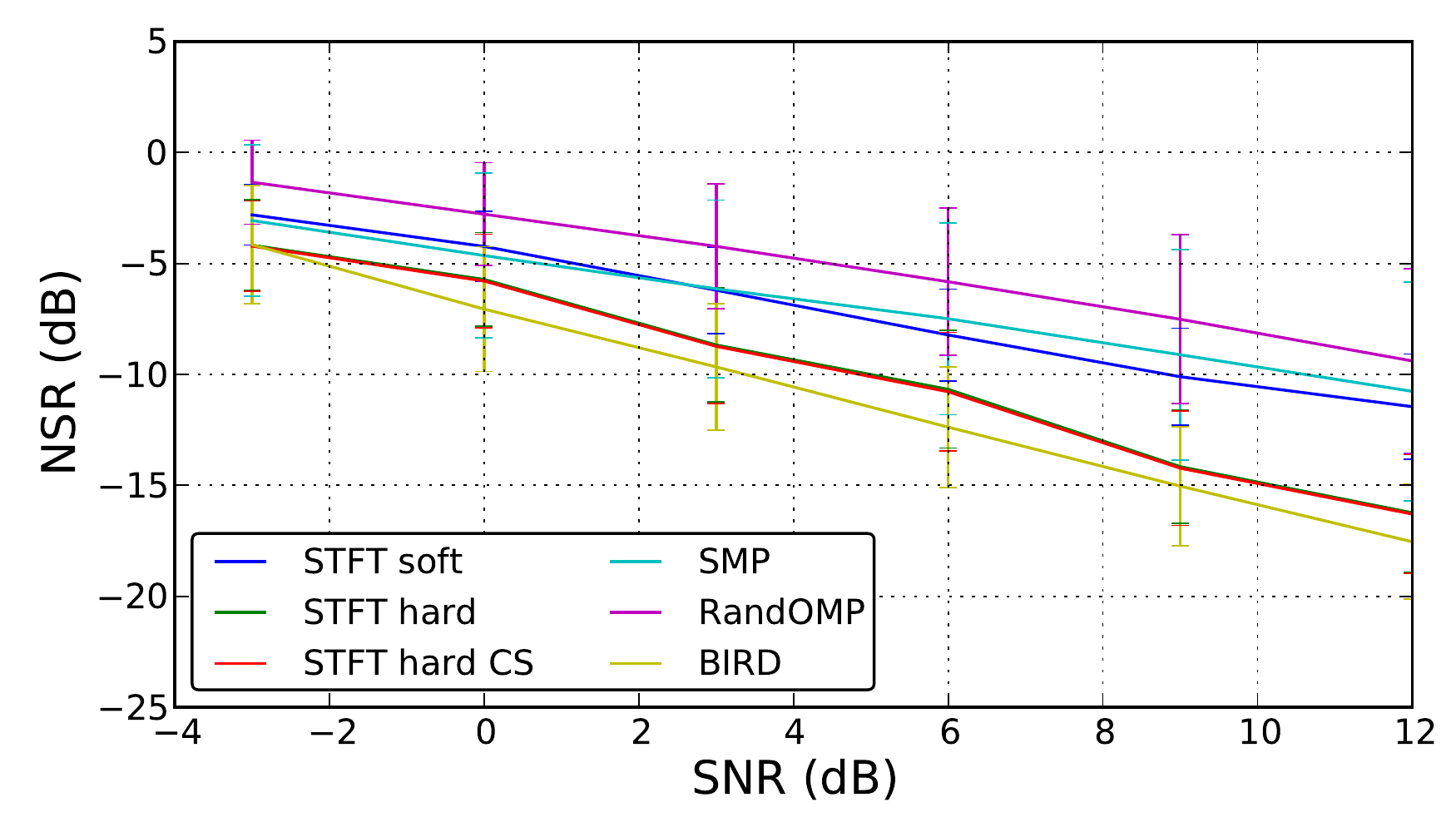}
\caption{ NMSE for various denoising methods on simulated single-sensor MEG data (score averaged over 20 trials) as a function of the noise level.}
\label{fig:single_simulated}
\end{figure}
Publicly available software~\cite{gramfort-etal:2014b} has been used to simulate MEG signals. A controlled level of white or colored noise $W$ (auto-regressive process fitted on real data) was added to a collection of $C$ smooth and oscillatory signals mimicking classical MEG evoked responses thanks to the use of a real forward solution. The recordings in the absence of noise are given in the form of a ground truth matrix $X$. Denoising methods can then be applied to $Y=X+W$ and compared using a Normalized Mean Squared Error (NMSE) ratio:
\begin{equation}
NMSE(\hat{Y}) = 10 \log_{10}\frac{\| X - \hat{Y}\|_F^{2}}{\|X\|_F^{2}} \enspace .
\label{NSR}
\end{equation}

Figure~\ref{fig:single_simulated} illustrates the performance of the proposed approach in terms of reconstruction for a single-sensor signal. BIRD outperforms competitive methods, even informed by the oracle noise level.

\begin{figure}
\centering
\includegraphics[width=8cm]{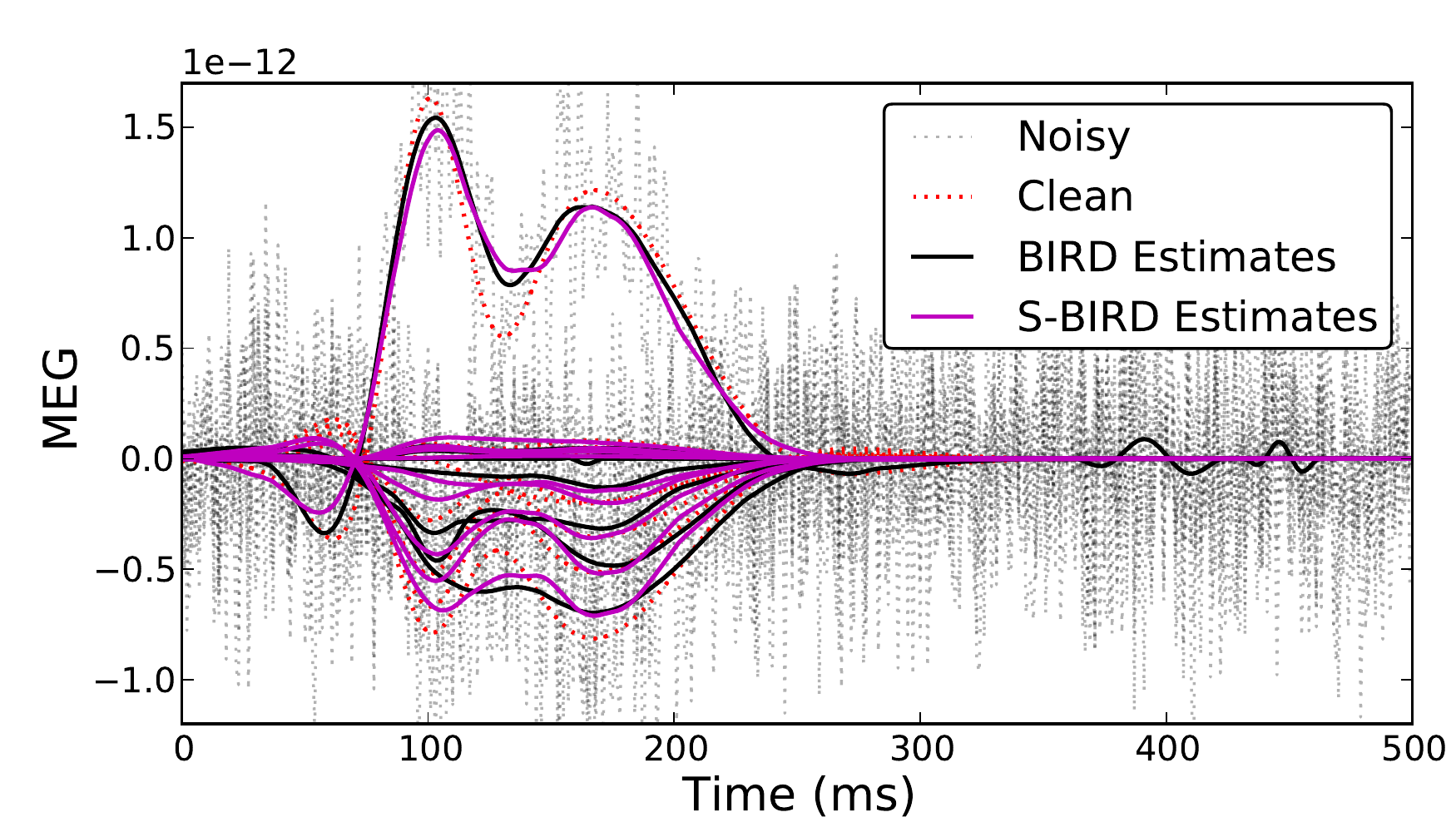}

\includegraphics[width=8cm]{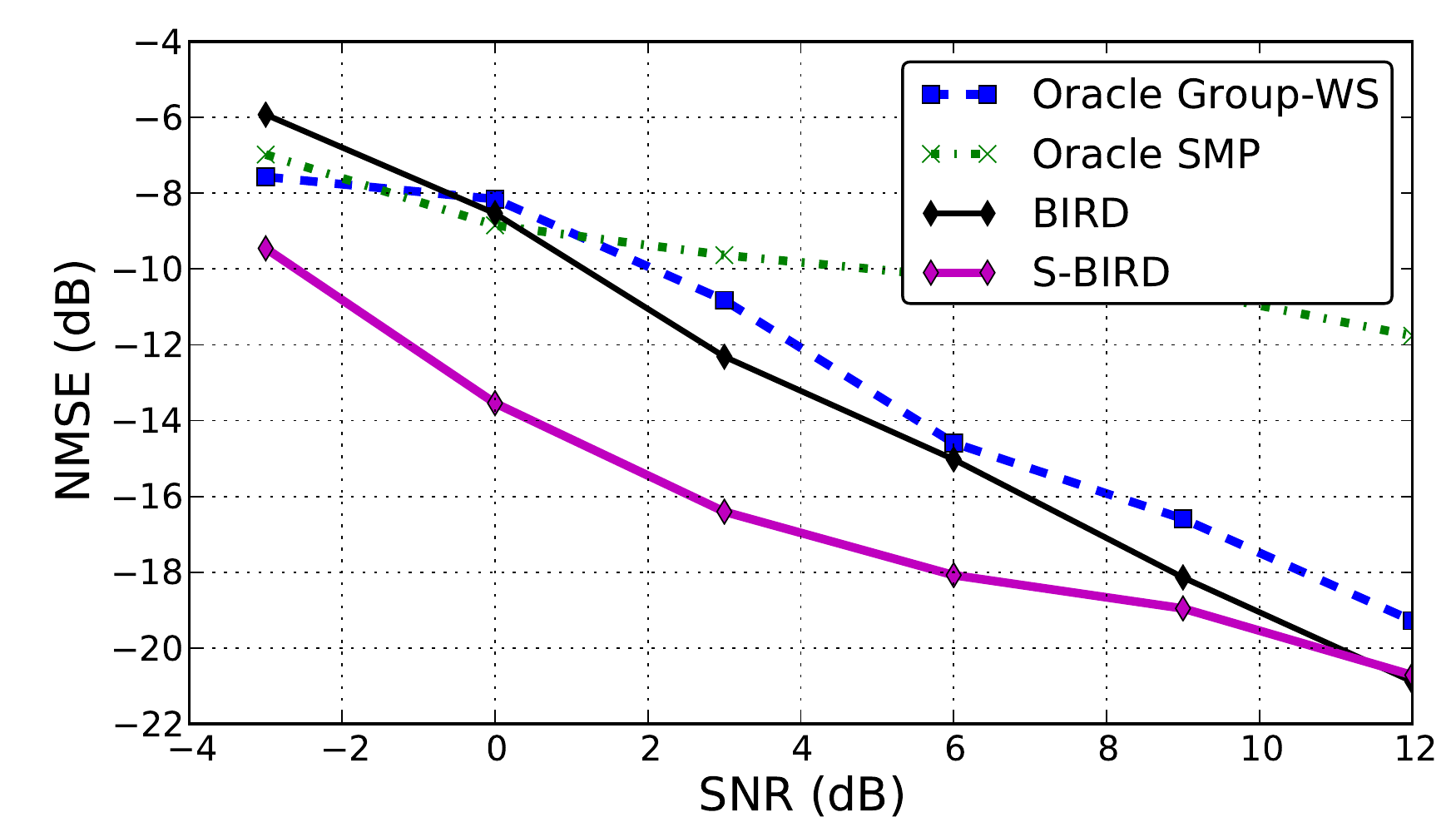}
\caption{Top: examples of BIRD and S-BIRD denoising for simulated multi-sensor MEG signals (evoked response corrupted by white noise). Bottom: NMSE (dB) for various methods.}

\label{fig:multi_simulated_white}
\end{figure}

Given the same test framework, we now evaluate the denoising capabilities on multichannel signals. We compare the performance of BIRD being applied independently to each sensor (\emph{i.e.} not taking any structure into account) to the Structured version S-BIRD. In this simulation study, all channels contain information and we set $l=1$.
For comparison, we use group soft thresholding methods using wavelets (Daubechies wavelets with 3 vanishing moments) and STFT, and present the best results obtained while varying the threshold parameter (labelled Oracle Group WS). Finally, it is compared to SMP applied independently on each sensor. As shown on Figure \ref{fig:multi_simulated_white}, by taking cross-sensor correlation into account at the tom selection level, S-BIRD improves over BIRD and outperforms other methods. This improvement is even more visible for colored noise and with real data (see supplementary material for additional figures).
\section{Conclusion}
We propose a greedy strategy that relies on averaging the results of
multiple runs of random sequential pursuits, each of which can select a
different number of atoms and reach a different approximation level
that is determined by a signal-independent stopping criterion. The only
parameter $p$ depends solely on the dictionary. 


The algorithm is fast as it avoids computing all projections while
using FFT-based MDCT or wavelet dictionaries.
An enhanced version S-BIRD, taking into account atom correlations between
multiple sensors achieves even better results on simulated data with white
or colored noise, as well as on MEG data. The multiple runs averaging strategy, also called \emph{bagging} \cite{Breiman1996} in the machine learning literature, reduces the estimation variance of a single pursuit, and is a key ingredient of the BIRD algorithm.

{
\clearpage

\bibliographystyle{ieeetr}
\bibliography{biblio}
}

\subsection*{Reproducing these results}
Python code to reproduce some of these figures and do further testing of BIRD and S-BIRD algorithms is freely accessible online: http://manuel.moussallam.net/birdcode
\subsection*{Coherence of the self-stopping criterion}
\begin{figure}[h!]
\centering
\includegraphics[width=12cm]{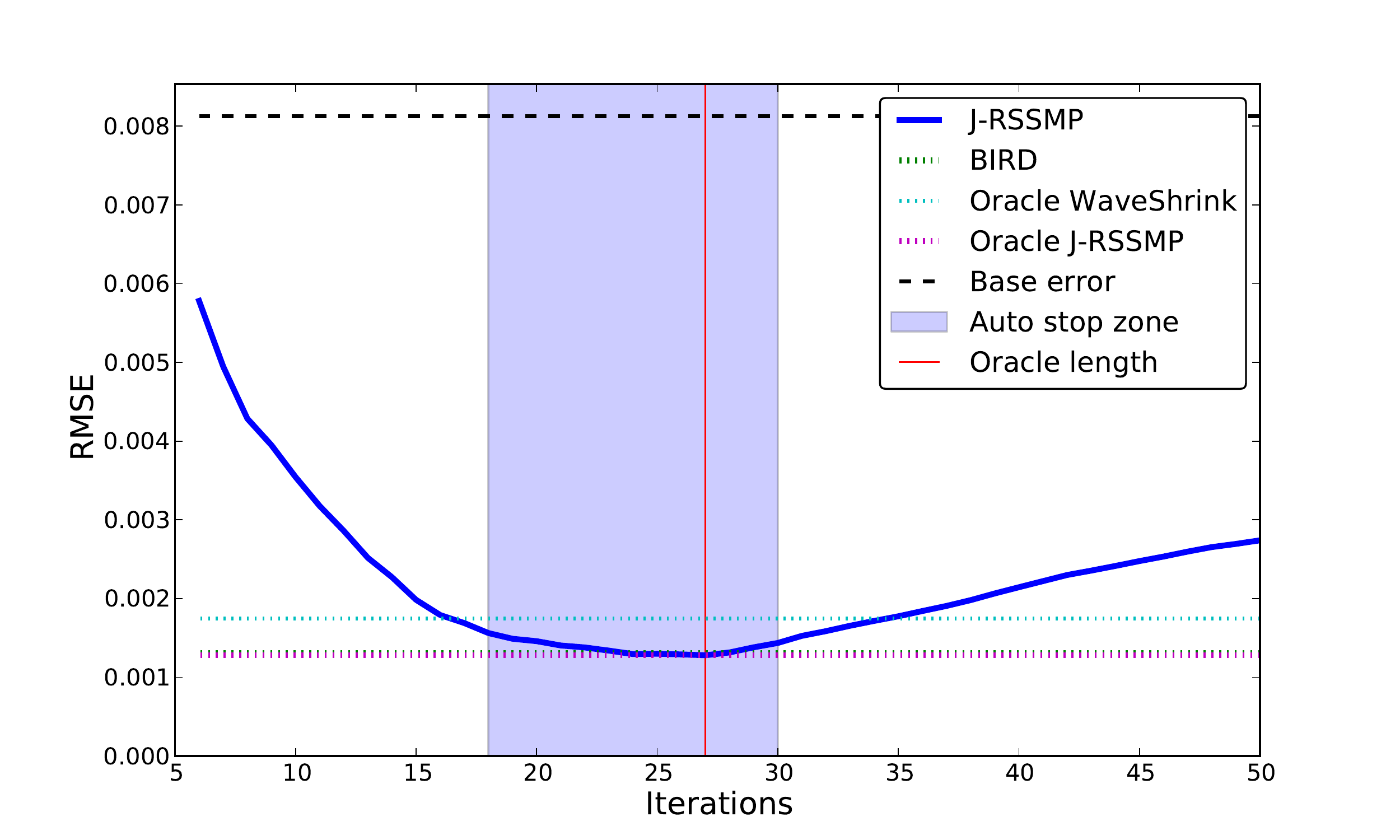}
\caption{Denoising performances of different algorithms on the synthetic \textit{Doppler} signal corrupted with an additive Gaussian white noise. The BIRD estimate is compared with the averaging of $J$ random pursuits (RSSMP). Each of the BIRD run stops somewhere in the blue zone. BIRD yields a comparable performance with the optimal RSSMP.}
\label{fig:coherence}
\end{figure}

BIRD runs $J$ decompositions in parallel that may each stop after a different number of iterations. For the synthetic Doppler signal, the self-stopping zone is represented in blue in Fig.~\ref{fig:coherence} (between 18 and 30 iterations). BIRD is compared with $J$ runs of the Random Sequential Subdictionary MP (RSSMP) algorithm for which the number of iterations is fixed knowing the noise level (\textit{Oracle} estimates). The error achieved by BIRD in the blind case matches the optimal oracle one. The same behavior can be observed with other synthetic signals, such as \textit{Blocks}, and other noise levels.

\clearpage
\subsection*{Setting the meta-parameter $p$}
\begin{figure}[h!]
\centering
\includegraphics[width=12cm]{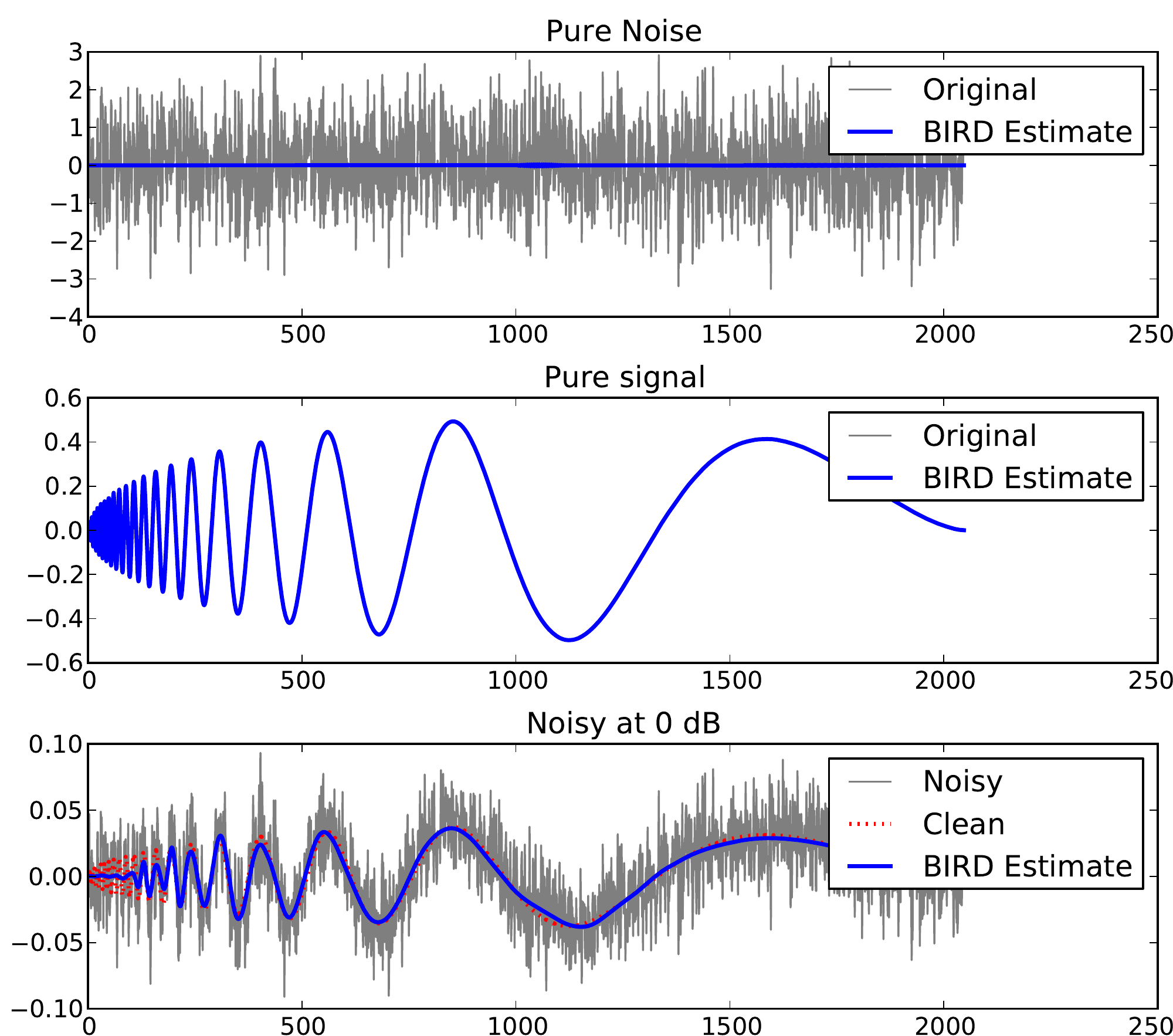}
\caption{Denoising results for (top) pure noise (middle) pure doppler signal and (bottom) noisy mixture with the same setting for $p$. The meta-parameter need not be adapted to each case. Reproduce using provided code.}
\label{fig:many_sigma}
\end{figure}

\begin{figure}[h!]
\centering
\includegraphics[width=16cm]{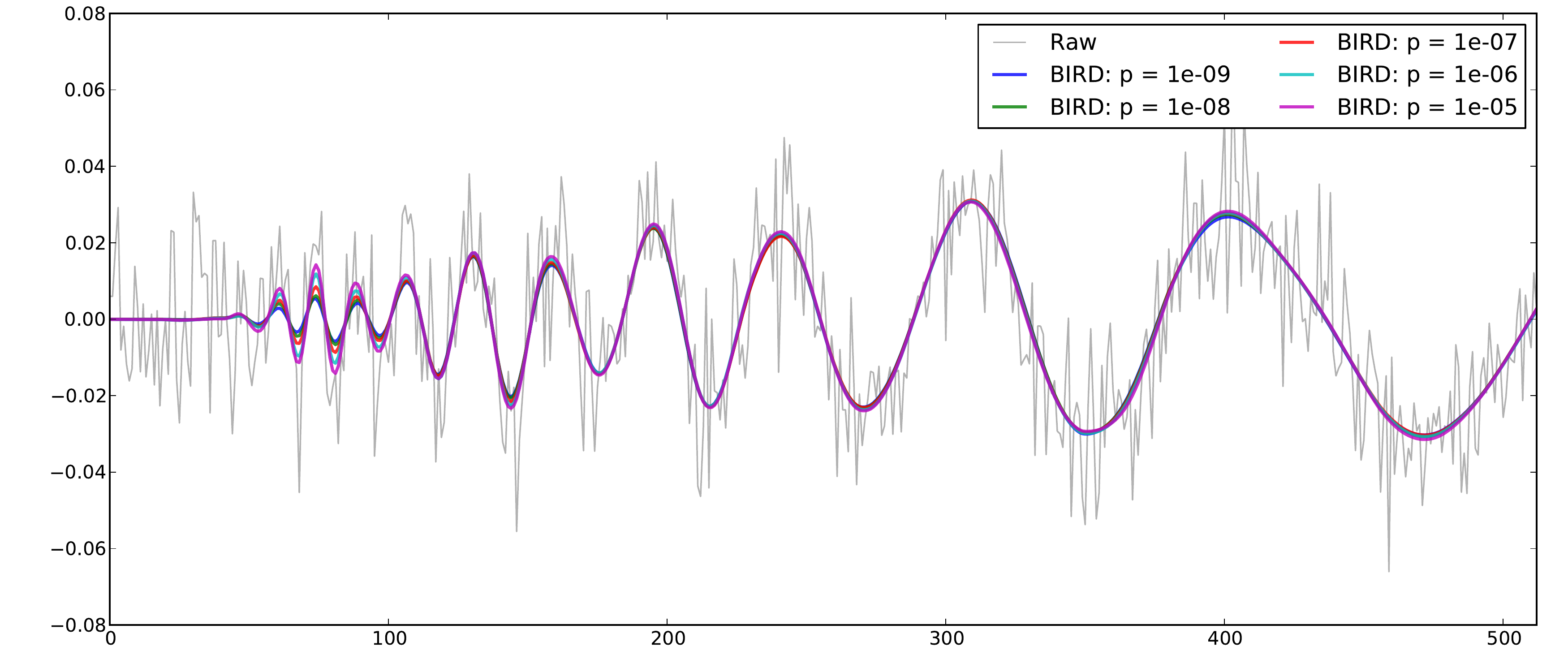}
\caption{Denoising results for noisy doppler (SNR 5dB) with various values of $p$. Many different orders of magnitude still lead to coherent estimates. Reproduce using provided code.}
\label{fig:many_p}
\end{figure}

The meta-parameter $p$ controls the number of iterations and the complexity of the estimates. However, it is different from the classical noise variance parameter $\sigma$. In fact, $p$ is quite independent from $\sigma$ as illustrated by Figure~\ref{fig:many_sigma}. Here we decompose three signals with exactly the same setting of $p$ (at a reference value of $1/M$). First signal is pure noise and the algorithm selects nothing. Second signal is a pure doppler corrupted by a white noise with a different variance and the algorithm selects atoms up to a very high reconstruction level. We report that intermediate noisy signals can be processed with $p$ left unchanged.
To demonstrate that the setting of the parameter $p$ is not critical, Figure~\ref{fig:many_p} shows denoising estimates with values of $p$ varying by multiple orders of magnitude. One can clearly see that the results are minimally impacted by the change of parameter $p$.

\subsection*{Complexity study on simulated MEG data}
\begin{figure}[h!]
\includegraphics[width=16cm]{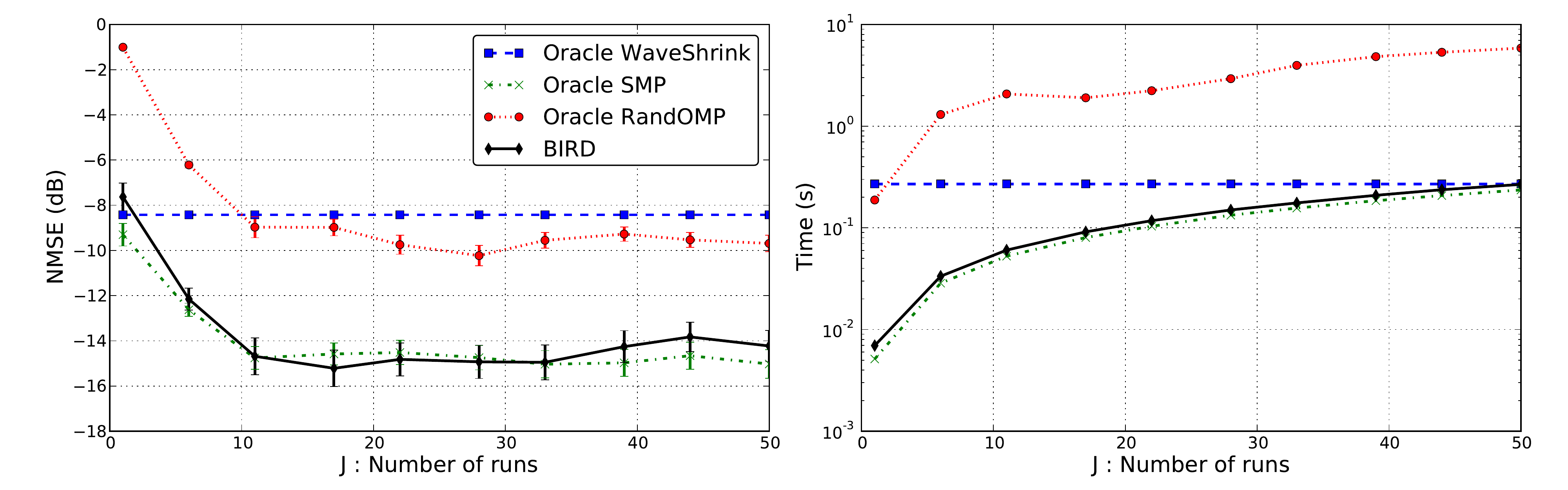}
\caption{Left: Denoising scores for various methods as a function of the number of runs. Right: Computing times. Simulated data with SNR of 12 dB. These scores reflect our own implementations of all these methods available at http://manuel.moussallam.net/birdcode (measured on a Intel Xeon CPU X5482 @ 3.20GHz × 4 with 32Gb of RAM).}
\label{fig:complexities}
\end{figure}

Results are reported in Figure~\ref{fig:complexities}.
With large signals and dictionaries, RandOMP has a prohibitive computational cost, not only due to the orthogonal update, but also for computing the projections of the signal onto the highly-redundant shift-invariant dictionary $\Phi$ (even when using fast transforms). It is also very sensitive to the input parameters, \emph{i.e.} the noise energy and the variance of the  sparse components.
Although it leads to higher complexity, WaveShrink is always combined with the cycle spinning method in this paper since higher performances are systematically obtains in this setting.
Increasing the number of runs J in our method increases computing time, however all the pursuits can be performed independently in parallel and the quality of the denoising (only evaluated through the NMSE here) quickly improves to reach a lower bound after approximately 15 decompositions.

\subsection*{Pink noise experiments}
\begin{figure}[h!]
\centering
\includegraphics[width=16cm]{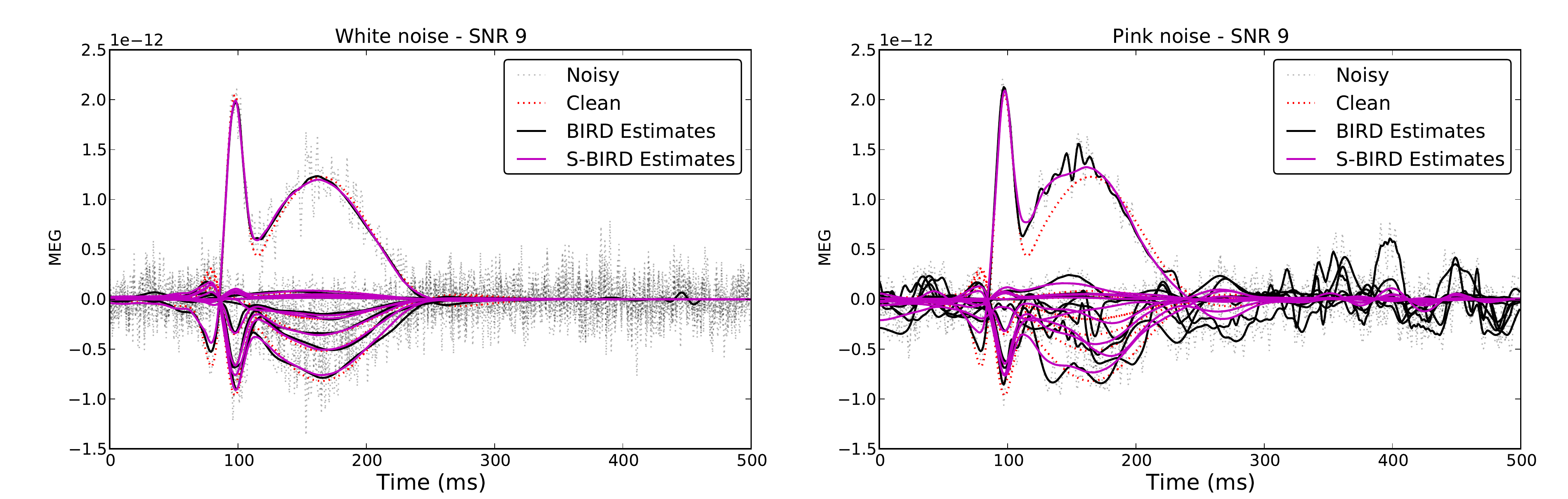}
\caption{Denoising of simulated MEG evoked signals with white (Left) and pink (Right) noise with BIRD and S-BIRD ($l=1$, $J=30$, $C=20$). SNR was set to 9dB and realistic brain sources were simulated with a first short early component at 100\,ms containing higher frequencies than a later second component peaking around 180\,ms.}
\label{fig:pink}
\end{figure}

\begin{figure}[ht]
\centering
\includegraphics[width=8cm]{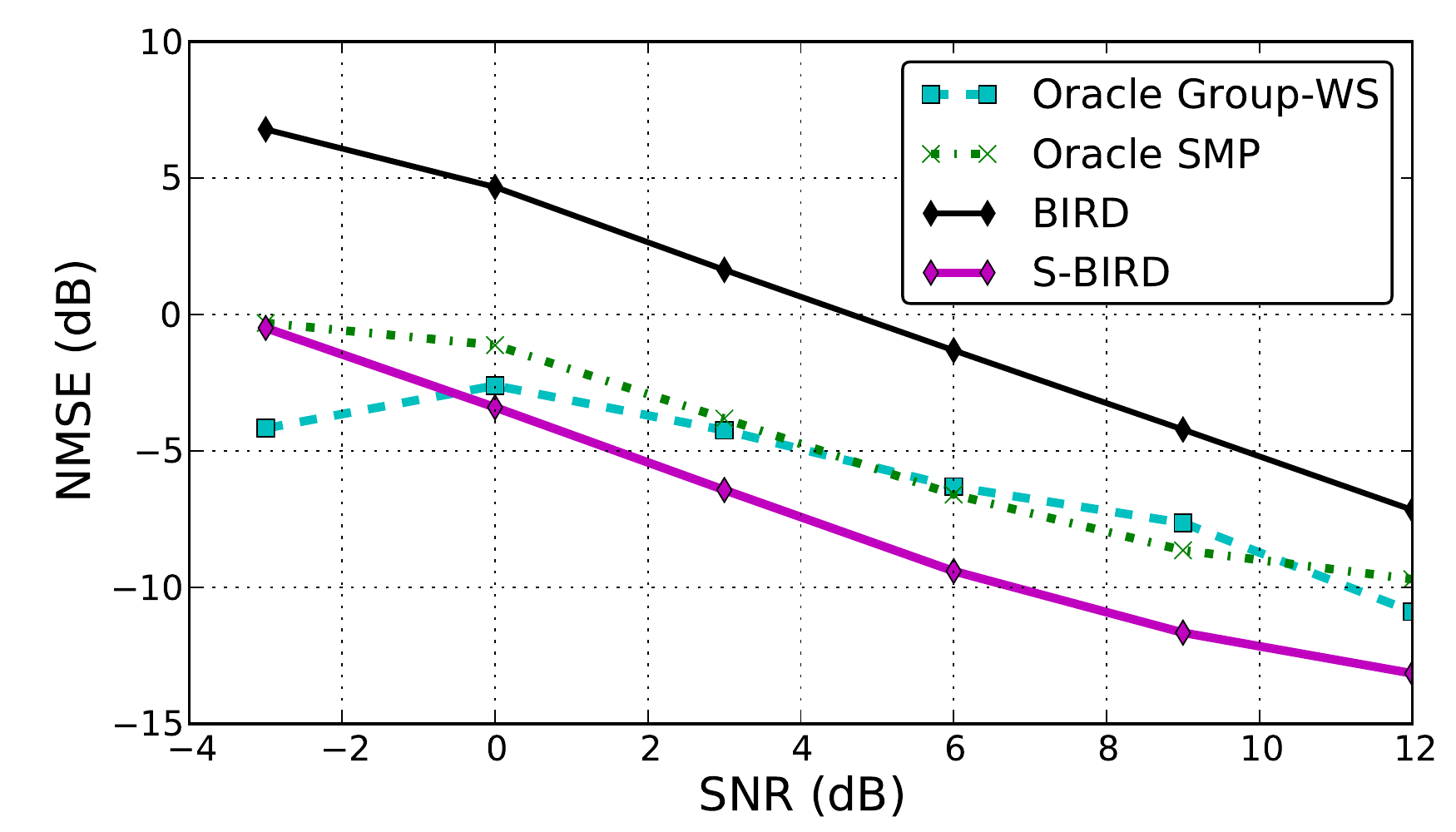}
\caption{NMSE obtained from various denoising methods when varying the level of a pink noise. The noise spectrum was estimated on real MEG signals using an autoregressive (AR) model of order 5. The structured version of BIRD gives very good results, especially at low SNRs while BIRD results deteriorate. }
\label{fig:multiple_simulated}
\end{figure}

The experiments on simulated MEG data can be extended to pink noise cases. The improvement brought by the structured model are even more visible (see Figures~\ref{fig:pink} and \ref{fig:multiple_simulated}). Indeed, the pink nature of the noise now violates the assumption of zero mean Gaussian distribution (GD) for the projections of the noise on the dictionary atoms, which causes BIRD to overfit the data. However, the structured version S-BIRD reduces this effect, mainly due to its ability to use interference across channels to improve atom selection.

\subsection*{Results on real MEG data}
We use data recorded with a Neuromag VectorView system (Elekta Oy, Helsinki, Finland) made available by the MNE software\footnote{http://martinos.org/mne/}. The signal used are from temporal and frontal planar gradiometers for a total of $C=78$ channels following an auditory stimulation in the left ear.
M/EEG data are typically the results of an experiment repeated $T$ times,
leading to $T$ measurement matrices $Y_t \in \mathbb{R}^{N \times C}$, $1 \leq t \leq T$.
Each repetition of the experiment is called a \emph{trial}.

In order to evaluate the quality of the denoised signals, the set of $T$ trials is separated in two sets of size $T_{learn}$ and $T_{test}$. The average matrix of the first set $Y_{learn} = \frac{1}{T_{learn}} \sum_{t=1}^{T_{learn}}Y_t$ is used to compute a denoised estimate $\hat{Y}_{learn}$. Assuming that each trial is corrupted by an independent noise signal, the matrix $Y_{test}$, obtained by averaging the $T_{test}$ left-out trials, can be used to evaluate the performance of the denoising procedure by quantifying how much $\hat{Y}_{learn}$ is close to $Y_{test}$ using an Averaged Noise-to-Signal Ratio (ANSR) measure:
\begin{equation}
ANSR(\hat{Y}_{learn}) = 10 \log_{10}\frac{\| Y_{test} - \hat{Y}_{learn}\|^2_F}{\|Y_{test}\|^2_F} \enspace .
\end{equation}

Table~\ref{tab:snsr} presents the results obtained on MEG data using the different methods. With only 5 trials, S-BIRD yields an estimate that is more reliable than a plain averaging of 30 trials. If one assumes that only about 80\% of the channels are simultaneously activated, then $l$ can be set to 0.8 which slightly improves S-BIRD performance.
For higher $T_{learn}$ values, we found threshold values that allowed group soft thresholding methods to slightly outperform S-BIRD. Their sensibility to this threshold is however important and it can not be chosen optimally in the absence of the test samples. The blind nature of our approach, in this real-life situation, is a clear advantage.
\begin{table}
\centering
\begin{tabular}{ | l | c | c | c | c |}
\hline
$T_{learn}$& Oracle Group WS & Oracle SMP & S-BIRD $l=1$ & S-BIRD $l=0.8$ \\
\hline 
\hline
 5  & -0.40 & -0.17 & -0.95 & \textbf{-0.99} \\
 10 & -1.15 & -1.03 & \textbf{-1.44} & \textbf{-1.44} \\
 15 & -1.47 & -1.28 & -1.51 & \textbf{-1.60} \\
 20 & -1.57 & -1.35 & -1.61 & \textbf{-1.62} \\
 30 & \textbf{-1.92 }& -1.51 & -1.79 & -1.83 \\
\hline
\end{tabular}
\caption{ANSR scores (dB) on real M/EEG data (78 channels). $T_{test}=25$}
\label{tab:snsr}
\end{table}

\begin{figure}
\centering
\includegraphics[width=16cm]{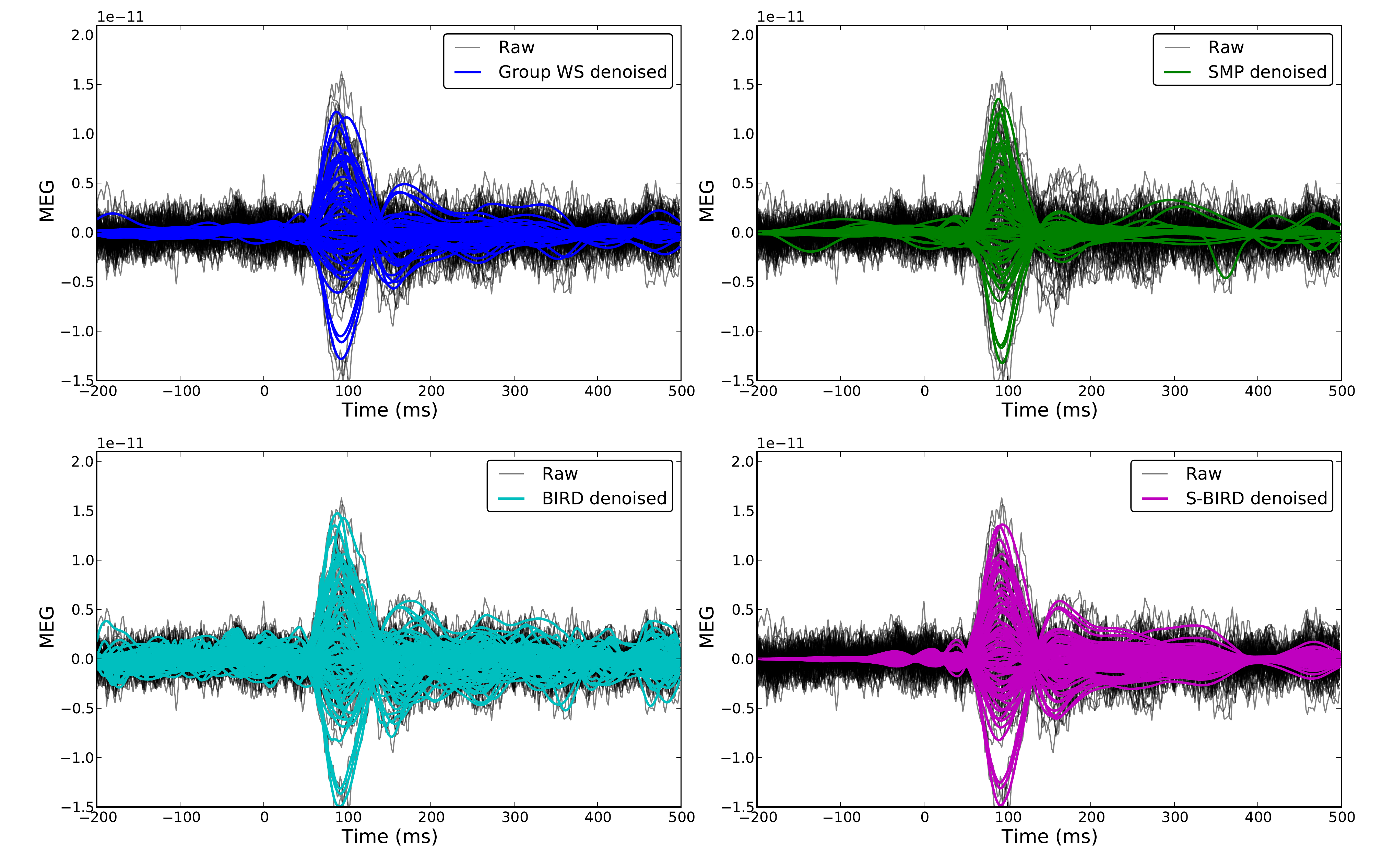}
\caption{Denoising of experimental MEG data acquired with a Neuromag VectorView system (Elekta Oy, Helsinki, Finland) from temporal and frontal planar gradiometers for a total of $C=78$ channels following an auditory stimulation in the left ear. Results presented are obtained by averaging 30 trials. The ability of S-BIRD to set almost to zero the signals prior to the stimulus onset at t=0 is striking. No amplitude bias is observed on the brain response peaking at around 100\,ms.}
\end{figure}

\end{document}